\documentclass{article}

\usepackage{PRIMEarxiv}

\usepackage[utf8]{inputenc} 
\usepackage[T1]{fontenc}    
\usepackage{hyperref}       
\usepackage{url}            
\usepackage{booktabs}       
\usepackage{amsfonts}       
\usepackage{nicefrac}       
\usepackage{microtype}      
\usepackage{lipsum}
\usepackage{caption}
\usepackage{fancyhdr}       
\usepackage{graphicx}       
\graphicspath{{media/}}     

\title{Expert Validation of Synthetic Cervical Spine Radiographs Generated with a Denoising Diffusion Probabilistic Model}

\author{
    \centering
    \begin{minipage}{0.95\textwidth}
    \centering
    \textbf{Austin A. Barr\textsuperscript{1\thanks{Correspondence: Austin A. Barr, austin.barr@ucalgary.ca}}}, 
    \textbf{Brij S. Karmur\textsuperscript{2}}, 
    \textbf{Anthony J. Winder\textsuperscript{3}}, 
    \textbf{Eddie Guo\textsuperscript{4}},
    \textbf{John T. Lysack\textsuperscript{3}}, 
    \textbf{James N. Scott\textsuperscript{3}}, 
    \textbf{William F. Morrish\textsuperscript{3}}, 
    \textbf{Muneer Eesa\textsuperscript{3}},
    \textbf{Morgan Willson\textsuperscript{3}}, 
    \textbf{David W. Cadotte\textsuperscript{2}}, 
    \textbf{Michael M.H. Yang\textsuperscript{2}}, 
    \textbf{Ian Y.M. Chan\textsuperscript{5}},
    \textbf{Sanju Lama\textsuperscript{6}}, 
    \textbf{Garnette R. Sutherland\textsuperscript{6}}\\[0.75em]
    \textnormal {\textsuperscript{1}Cumming School of Medicine, University of Calgary; 
    \textsuperscript{2}Division of Neurosurgery, Department of Clinical Neurosciences, University of Calgary; \textsuperscript{3}Department of Radiology, University of Calgary; \textsuperscript{4}Division of Neurosurgery, Department of Surgery, University of Toronto; \textsuperscript{5}Department of Medical Imaging, University of Toronto; \textsuperscript{6}Project neuroArm, Department of Clinical Sciences, University of Calgary}
    \end{minipage}
}

\begin{document}
\maketitle

\begin{abstract}
\textbf{Background}:  Application of machine learning (ML) in neurosurgery is often constrained by the difficulty of assembling, sharing, and utilizing large, high-quality imaging datasets. Synthetic data offers a novel solution to this challenge by enabling the creation of privacy-preserving images, which can be generated at scale.\\
\\
\textbf{Objective}: To evaluate the feasibility of using a denoising diffusion probabilistic model (DDPM) to generate realistic synthetic lateral cervical spine radiographs.\\
\\
\textbf{Methods}: A DDPM was trained on the Cervical Spine X-ray Atlas (CSXA; 4,963 lateral radiographs), and monitored using training/validation loss and Fréchet inception distance (FID) to quantify model convergence and synthetic image realism. Blinded expert validation (“clinical Turing test”) was conducted with six neuroradiologists and two spine fellowship-trained neurosurgeons. Each expert reviewed 50 image quartets containing one ground truth radiograph and three synthetic images derived from separate training checkpoints. For each quartet, experts were tasked with identifying the real image and rating each image’s realism on a 4-point Likert scale (1 = unrealistic, 4 = fully realistic). Images were also evaluated for potential training-data memorization via a nearest-neighbor search implemented with \textit{imagededup}, using vision transformer embeddings and ranking real-synthetic pairs by cosine similarity.\\
\\
\textbf{Results}: Experts correctly identified the real image in 29.0\% of quartet trials, with low inter-rater agreement (Fleiss’ $\kappa$ = 0.061). Mean realism scores were 3.323 for real images and 3.228, 3.258, and 3.320 for synthetic images from the three checkpoints. Paired Wilcoxon signed-rank tests showed no significant differences between real and synthetic mean ratings (unadjusted \textit{p} = 0.128, 0.236, 1.000; Holm-adjusted \textit{p} = 0.383, 0.471, 1.000; two-sided). No visually explicit memorization was identified among the nearest-neighbor pairs. We also include a large-scale dataset of 20,063 synthetic radiographs.\\
\\
\textbf{Conclusions}: We present the generation of synthetic cervical spine X-ray images that are statistically indistinguishable in realism and quality from real radiographs in blinded expert review. This novel application of DDPM highlights the potential to generate large-scale neuroimaging datasets to support ML model training for landmarking, segmentation, and classification tasks.

\end{abstract}


\section{Introduction}

Machine learning (ML) is increasingly integrated into neurosurgical and neuroimaging research; however, progress can be constrained by the limited availability of large-scale training data \cite{Ioannidis2020, Chan2025}. Patient volumes are often limited \cite{Zimmermann2019, Zlochower2020}, institutional data are heterogeneous and frequently incomplete \cite{Newgard2015}, and the labor required to curate, standardize, and de-identify raw data is substantial \cite{Willemink2020}. Privacy and regulatory constraints further complicate data sharing \cite{Bentzen2021, Legido2025, Ness2007}, which may otherwise be used to compile multi-institutional datasets to address the limitations of single-center studies and improve model performance and generalizability. \\
\\
Synthetic data has emerged as a pragmatic complement to real-world data (RWD) under these constraints. Synthetic data that are not tied to real patient records but preserve the salient properties or statistical structure of RWD can be produced at scale and shared without regulatory restrictions \cite{Bellovin2019, Rajotte2022}. Increasingly, synthetic tabular data has demonstrated significant fidelity to perioperative RWD \cite{BarrJNS2025, BarrFrontiers2025, Davalan2025}. For example, synthetic tabular neurosurgical data was used to amplify small RWD sample sizes, train ML models to predict patient outcomes, and conduct analyses with comparable results to ground truth clinical findings \cite{BarrJNS2025}. Synthetic imaging is advancing along a similar trajectory \cite{Koetzier2024}, albeit at a slower pace, due to the added complexity of achieving visual fidelity. One method used to generate synthetic images is a generative adversarial network (GAN), which employs two competing networks to improve image fidelity \cite{Goodfellow2014}. Early evaluations of GAN-based synthetic imaging data have demonstrated significant fidelity in expert reviews \cite{Jang2023, Myong2023}. Much of this work, however, has been concentrated on chest radiography, where datasets such as MIMIC-CXR (377,110 images) \cite{Johnson2019} and CheXpert (224,316 images) \cite{Irvin2019} enable large-scale training and evaluation. In contrast, neuroimaging datasets are typically much more limited in both size and availability \cite{Ioannidis2020, Chan2025}. This ‘abundance irony’ underscores why neurosurgical and neuroimaging research stands to benefit disproportionately from synthetic data.\\
\\
Another method of synthetic image generation is the denoising diffusion probabilistic model (DDPM) \cite{Ho2020}. The DDPM generates images by iteratively denoising random noise, producing high-quality and diverse synthetic images. Emerging evidence suggests that DDPM-generated images exhibit superior image quality and fidelity to those produced by GANs \cite{Dhariwal2021}. Yet, the application of a DDPM to generate new samples of synthetic neuroimaging data, and rigorous assessment of image quality and realism remains largely unexplored.\\
\\
In this study, we investigate whether a DDPM can generate realistic synthetic lateral cervical spine X-rays, when trained on the open-access Cervical Spine X-ray Atlas (CSXA) of 4,963 radiographs \cite{Ran2024, RanPaper2024}. We conducted a multi-expert blinded validation (“clinical Turing test”) in which neuroradiologists and spine fellowship-trained neurosurgeons were tasked with discriminating between real and synthetic images. We hypothesized that experts would be unable to distinguish real from synthetic images.

\section{Methods}

\subsection{Real-World Data Source and Data Pre-Processing}

Training images were drawn from the CSXA, an open-access dataset of 4,963 lateral cervical spine radiographs \cite{Ran2024, RanPaper2024}. CSXA was selected for its scale, open availability, and accompanying metadata and annotations for downstream ML tasks (i.e., vertebral keypoints). All CSXA radiographs were included and resampled to 256x256 pixels. Negative scans (i.e., cortical bone rendered dark) were intensity-inverted to homogenize the coloration of training data. 

\subsection{DDPM Implementation and Training}

We implemented a DDPM training and sampling setup with Python code adapted from the serag‑ai “I‑SynMed” repository \cite{SynMed, Hosseini2025}. To optimize the hyperparameters of the model training session, we split the CSXA dataset into training and validation partitions comprising 4219 (85\%) and 744 (15\%) images, respectively, by random sampling. Hyperparameters included batch size (8), learning rate (5e-5), and the number of training steps (up to 160,000). Validation loss was monitored at ‘model checkpoints’, saved every 2,000 training steps.\\
\\
To generate the final model used for sampling, a separate instance of the DDPM was trained using 100\% of the available images to maximize data utilization. During this final training session, distributional similarity between real and synthetic images was evaluated using Fréchet inception distance (FID), which compares the statistical properties of feature representations (computed with a pretrained Inception network) between the two sets of images \cite{Heusel2017, Hosseini2025}. FID scores were computed between the 4,963 real images and 7,250 synthetic images generated at each interval of 20,000 training steps.\\
\\
Training and sampling were performed on institutional NVIDIA A100 or V100 GPUs.

\subsection{Synthetic Dataset Preparation and Post-Processing}

Synthetic image sets (1,000 images each) were generated using the DDPM at three specific model checkpoints, determined from the loss curve trajectories and FID scores. Following generation, synthetic images were standardized to left-facing orientation to ensure consistent anatomical alignment. A manual review was then conducted to exclude any samples with implausible anatomy (e.g., pharynx and trachea posterior to the spinal column, spinal column anterior and posterior elements inverted, etc.). From each of the three synthetic sets, 50 images were randomly selected for expert validation, along with 50 randomly selected real scans from CSXA. For the validation task, images were grouped into quartets: each containing one real image and three synthetic images (one from each checkpoint).\\
\\
In parallel, we generated larger synthetic image sets from each checkpoint and pooled them into a single, large-scale dataset intended for publication. Each large-scale sampling job was conducted up to a maximum institutional runtime limit of 24 hours. The generated sets were manually reviewed using the same post-generation screen for gross abnormalities.

\subsection{Memorization Audit}

To ensure that the model was not simply memorizing training data, we conducted a near-duplicate search of all generated images using the open-source \textit{imagededeup} library \cite{Imagededup}, a commonly used library for image similarity and duplicate detection \cite{Kaur2024, Zhang2024}. We computed vision transformer (ViT) embeddings for all real and synthetic images, ranked nearest neighbors by cosine similarity, and extracted the top 100 most similar real-synthetic pairs. Two authors (AAB and BSK) independently performed side-by-side visual review of these 100 pairs, adjudicating for explicit memorization. \textit{A priori}, we defined explicit memorization as near-identical vertebral contours, soft-tissue silhouettes, or acquisition artifacts beyond plausible sampling variability. 

\subsection{Blinded Expert Validation}

Eight experts (six neuroradiologists and two spine fellowship-trained neurosurgeons) from the University of Calgary and the University of Toronto completed a blinded validation task using the Qualtrics survey platform (Qualtrics, Provo, UT). Each expert reviewed 50 quartets (presented sequentially on separate pages), containing one real CSXA radiograph and three synthetic images sampled independently from the three checkpoints. For every quartet, raters identified the image they believed was real and assigned subjective realism ratings to each image on a four-point Likert scale (1 = unrealistic; 4 = fully realistic). 

\subsection{Statistical Analysis}
\setcounter{footnote}{0}
Identification accuracy was computed as the proportion of quartets in which the real image was correctly identified, aggregated across raters. Inter-rater agreement was quantified with Fleiss’ $\kappa$ \cite{Fleiss1971}. For realism ratings, each rater’s mean score per image group was computed. The mean score for the real images was then compared with the mean score for each of the three synthetic image groups using paired, two-sided Wilcoxon signed-rank tests. Holm’s procedure was applied to adjust for multiple comparisons. Statistical analyses were performed in R (version 4.4.2, The R Project for Statistical Computing) and rating distributions were visualized with violin plots created using the Python library Matplotlib\footnote{https://matplotlib.org}.

\subsection{Ethical Considerations}

The CSXA is an open-access dataset released under Creative Commons Attribution 4.0 International (CC BY 4.0) \cite{Ran2024}. We have made our large-scale pooled dataset available for inclusion with this publication, consistent with principles of open science and the permissive terms of the CC BY 4.0 license. As members of the study team conducted a blinded review of images and we did not include human participants beyond co-authors, formal research ethics board approval was not required.

\section{Results}

\begin{figure}[!h]
  \centering
  \includegraphics[width=0.6\textwidth]{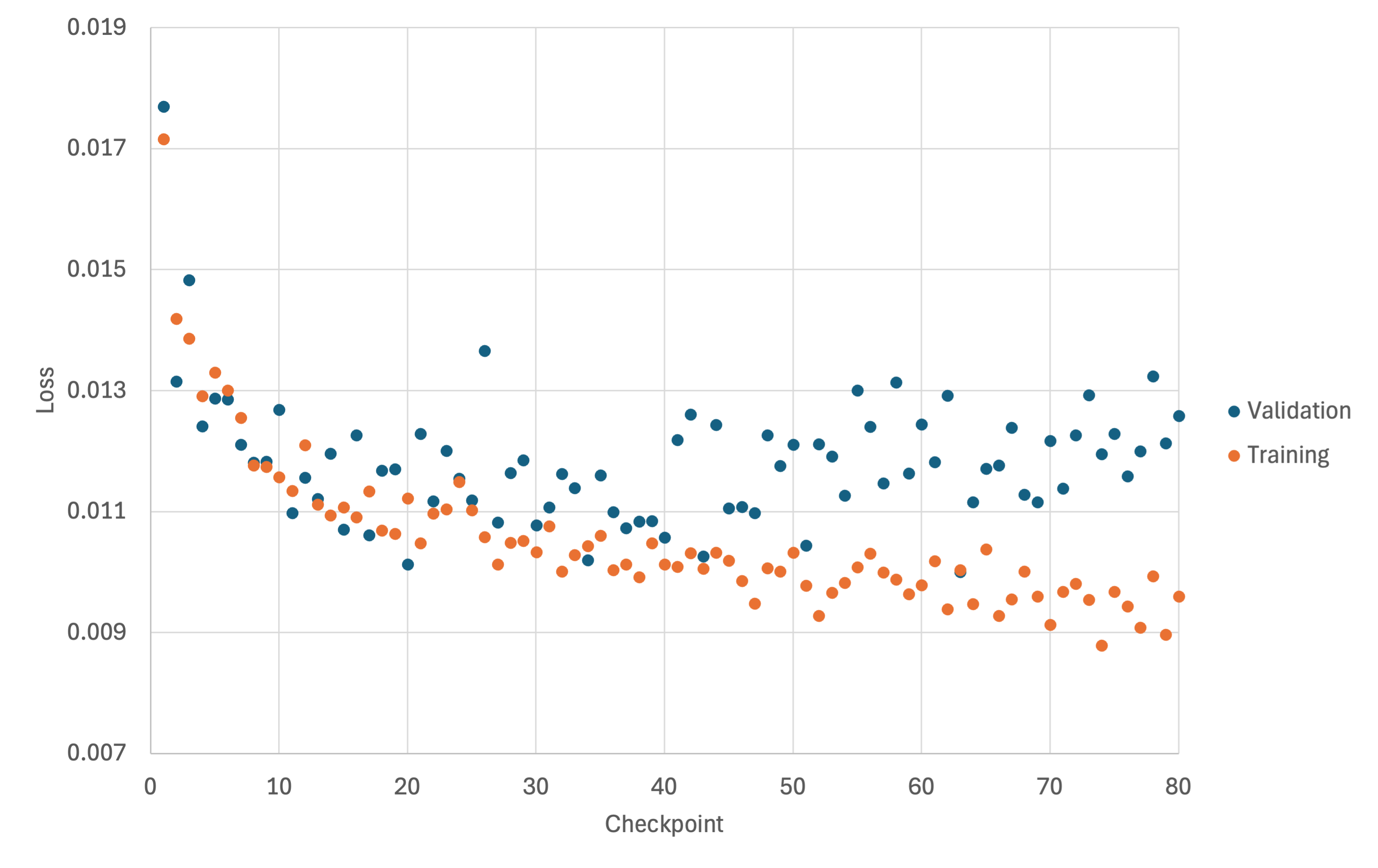}
  \captionsetup{width=0.8\textwidth}
  \caption{Training and Validation Loss of the Denoising Diffusion Probabilistic Model (DDPM). A separate instance of the DDPM was trained from a random sample comprising 85\% of the available training data, and the remaining 15\% was used for validation. The model loss was computed for the training and validation data partitions at each model checkpoint (every 2,000 training steps).}
  \label{fig:figure1}
\end{figure}

\begin{figure}[!h]
  \centering
  \includegraphics[width=0.6\textwidth]{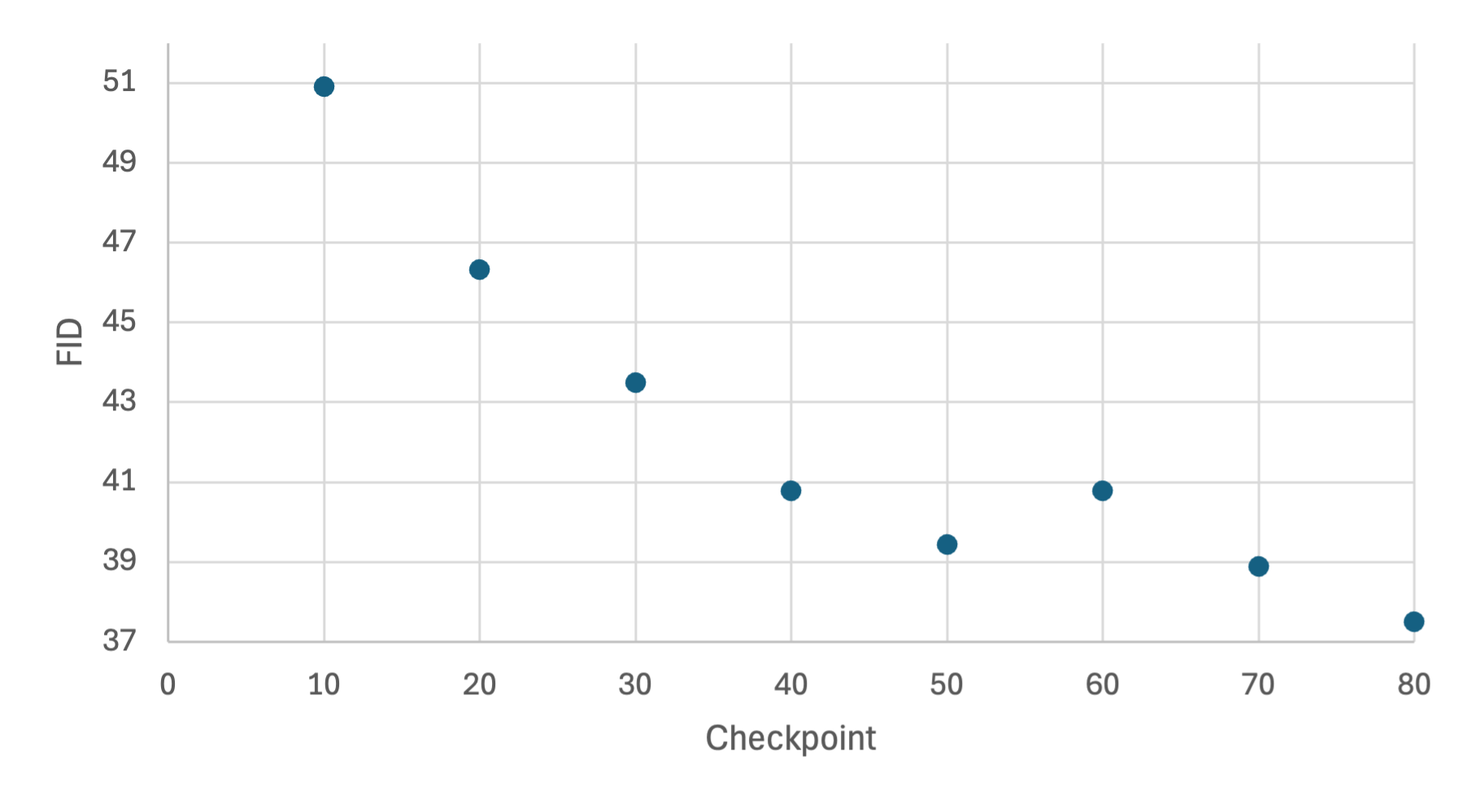}
  \captionsetup{width=0.8\textwidth}
  \caption{Fréchet Inception Distance (FID) Across Training Checkpoints. FID was computed between the 4,963 real clinical images and a set of 7,250 synthetic images generated at every 10th model checkpoint (once every 20,000 training steps).}
  \label{fig:figure2}
\end{figure}

Training and validation losses converged between checkpoints 30 and 40, diverging thereafter (Figure \ref{fig:figure1}). FID computed from 7,250-image synthetic batches declined over training (Figure \ref{fig:figure2}), indicating ongoing improvement in sample quality despite the loss divergence. On this basis, we selected checkpoints 34 and 40 to capture the period of loss convergence, and checkpoint 80 to assess potential benefits from extended training time not reflected in the loss metrics.\\
\\

\begin{figure}
  \centering
  \includegraphics[width=0.8\textwidth]{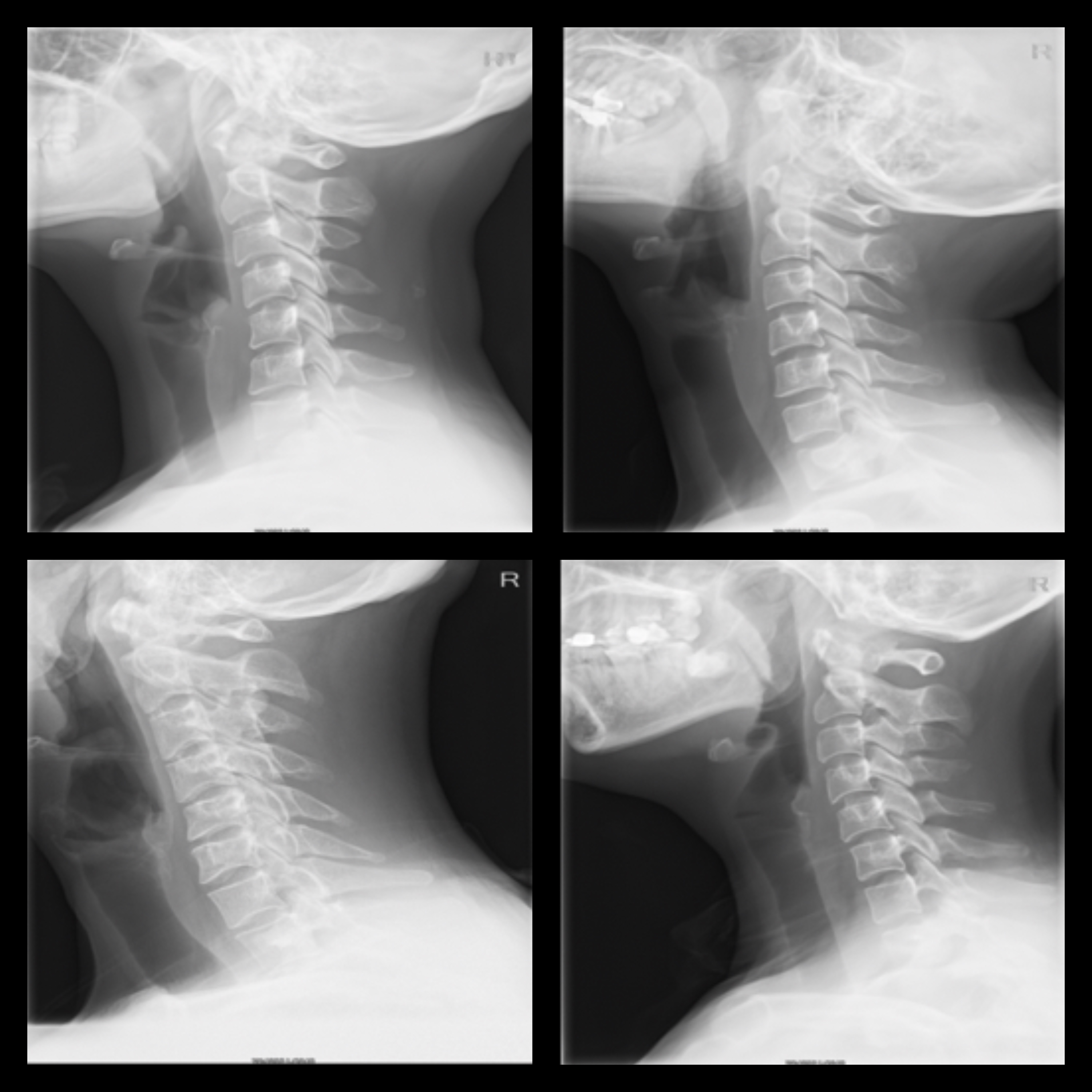}
  \captionsetup{width=0.8\textwidth}
  \caption{Sample Quartet of Real and Synthetic Lateral Cervical Spine X-Ray Images. One of 50 quartets presented to neuroradiologists and spine fellowship-trained neurosurgeons for evaluation. The figure includes one real patient’s X-ray (bottom left) and three synthetic images generated from saved model weight checkpoints: 34 (top left), 40 (bottom right), 80 (top right), where each checkpoint corresponds to 2,000 training steps.}
  \label{fig:figure3}
\end{figure}

\begin{figure}
  \centering
  \includegraphics[width=0.8\textwidth]{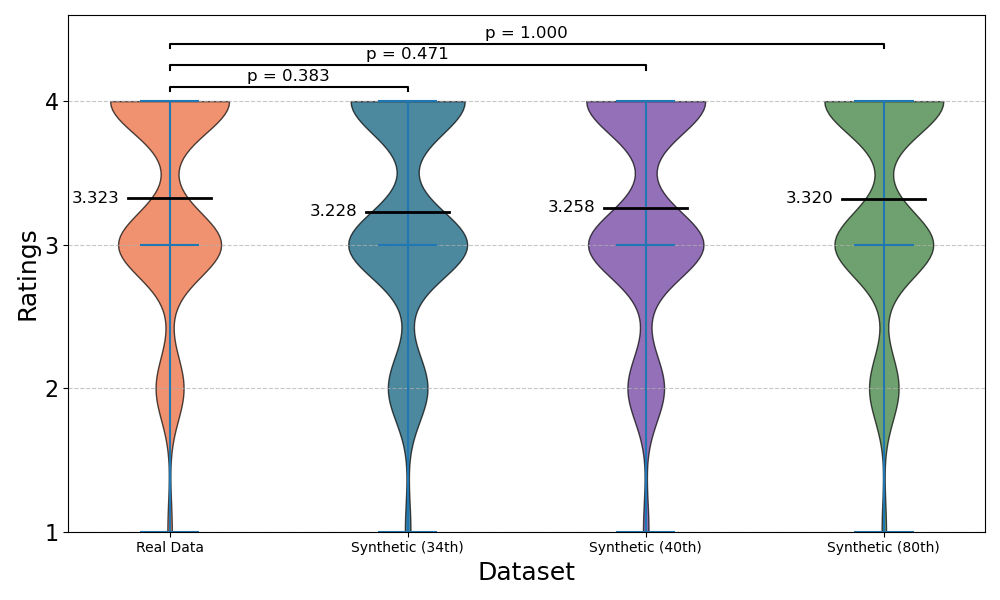}
  \captionsetup{width=0.8\textwidth}
  \caption{Comparison of Image Realism Ratings Across Real and Synthetic Datasets. Violin plots display the distribution of ratings for the real data and three synthetic datasets generated from saved model weight checkpoints: 34, 40, 80, where each checkpoint corresponds to 2,000 training steps. \textit{p}-values are displayed with Holm correction (unadjusted \textit{p} = 0.128, 0.236, 1.000).}
  \label{fig:figure4}
\end{figure}

Across 400 quartet judgements (eight raters with 50 quartets per rater), experts identified the real image in 29.0\% of trials. A sample quartet is provided in Figure \ref{fig:figure3}. Inter-rater agreement for the identification task was low (Fleiss’ $\kappa$ = 0.061). Mean realism ratings were 3.323 for real images and 3.228, 3.258, and 3.320 for synthetic images sampled from checkpoints 34, 40, and 80, respectively (Figure \ref{fig:figure4}). Paired Wilcoxon signed-rank tests showed no significant differences between real and synthetic mean ratings (unadjusted \textit{p} = 0.128, 0.236, 1.000; Holm-adjusted \textit{p} = 0.383, 0.471, 1.000; two-sided).\\

The memorization audit using ViT embeddings produced 100 real-synthetic pairs with the highest cosine similarity. Two independent reviewers found no visually explicit memorization among these top-similar pairs under the \textit{a priori} criteria. The top 100 most similar real-synthetic pairs are available at \footnote{\label{fn:dataset}https://doi.org/10.6084/m9.figshare.30436465.v1}.\\

From the 27,000 generated images, scans with gross anatomical abnormalities were removed. The resulting dataset, comprising 20,063 images from the three checkpoints, is available at \ref{fn:dataset}. A greater proportion of images were removed from the earlier checkpoints, reflecting a higher incidence of gross abnormalities in images sampled earlier in DDPM training. The specific counts of excluded images from each checkpoint are detailed in Table \ref{tab:supplementary_S1}.

\section{Discussion}

In this study, we presented a “clinical Turing test” for synthetic lateral spine radiographs generated by a DDPM trained on an open-access cervical spine X-ray dataset. The DDPM generated synthetic neuroimaging data that neuroradiologists and spine fellowship-trained neurosurgeons could not reliably distinguish from real clinical images. Identification accuracy was near chance, inter-rater agreement on real-image selection was low, and subjective realism ratings did not differ significantly between real and synthetic images. These findings complement quantitative indicators such as FID by grounding evaluation in domain-expert perception of anatomical plausibility and realism.\\

Anatomically faithful synthetic neuroimaging data offer a practical solution to long-standing RWD limitations. Neurosurgical patient volumes are often small and unevenly distributed across sites, constraining the assembly of large, representative training sets \cite{Ioannidis2020, Chan2025, Zimmermann2019, Zlochower2020}. By contrast, once a DDPM is trained, synthetic images can be produced at scale to enable greater availability of open-access datasets for academic research and ML model development. Clinical data are also heterogeneous and frequently incomplete \cite{Newgard2015, Willemink2020}, whereas synthetic images are generated in standard form and can augment dataset sizes. The labor necessary to de-identify RWD \cite{Willemink2020} and coordinate data-sharing agreements is significant \cite{Bentzen2021, Legido2025, Ness2007}. Synthetic images, which are not tied to specific individuals, can reduce de-identification burden and lower the privacy risks of sharing data between sites. Synthetic datasets may be generated from institutional data and shared in place of RWD to support large-scale, multi-institutional dataset curation. Synthetic datasets may also be generated to preserve the informational value of clinical data beyond the confines of data retention periods. Beyond research and ML model development, shareable datasets may provide educational opportunities to create large-scale teaching banks for trainees to more readily practice radiograph interpretation. In addition to trainee education, synthetic images can provide examples of pathological and non-pathological images for patient education (e.g., when demonstrating the difference between lateral cervical spine X-rays of a degenerative and a normal spine). Our release of a pooled synthetic lateral cervical spine dataset aims to catalyze these use cases \cite{BarrDataset}.\\

This work adds to a growing body of evidence evaluating the fidelity, utility, and privacy of synthetic data. Similar to past evaluations of GAN- and DDPM-based chest radiograph images \cite{Jang2023, Myong2023}, our evaluation of DDPM-generated spine X-rays demonstrated high expert-perceived realism under blinded review. Our present work expands on this literature in three ways. First, while chest X-ray pipelines have been trained on datasets numbering in the hundreds of thousands (e.g., MIMIC-CXR >370k; CheXpert >220k) \cite{Johnson2019, Irvin2019}, neuroimaging datasets are typically several orders of magnitude smaller \cite{Ioannidis2020, Chan2025}. Accordingly, we achieved comparable expert-perceived realism despite training on a much smaller corpus (<5k lateral radiographs). Importantly, our top-k nearest neighbor audit found no visually explicit reproduction, mitigating concerns that high realism merely reflects memorization or overfitting under a small training dataset approach. Second, we broadened our clinical Turing task to include a multi-expert evaluation, beyond a radiologist-only evaluation. Third, we used a head-to-head design that allowed experts to directly compare one real image with three synthetic images, rather than presenting images in isolation. This comparative format may have provided subspecialists with an additional opportunity to detect subtle inconsistencies across images and increase sensitivity to differences that may not be readily apparent in isolation. Despite the altered setup, we observed low accuracy and inter-rater agreement on the identification task.\\

Beyond situating our results within the landscape of synthetic imaging data, it is instructive to note parallel progress in synthetic tabular neurosurgical data. Recent work has demonstrated that synthetic cohorts can preserve outcome-relevant signals for real-world analyses and model training \cite{BarrJNS2025}. Building on both lines of evidence, the next step for our imaging-data pipeline is to adopt similar, task-based utility benchmarks that demonstrate concrete downstream value to our dataset. To that end, follow-up work will train models on our synthetic radiographs (alone and in mixed training sets with RWD) for relevant tasks (e.g., vertebral landmarking, anatomical segmentation, disease classification) and compare performance with baseline models trained exclusively on RWD.

\subsection{Limitations}

All CSXA radiographs were resampled to 256x256 pixels to simplify modeling and standardize training, which can distort anatomical proportions. Future work will scale training and synthesis to higher resolutions. Initial evaluation of similarity between the distributions of real and synthetic images with FID is not a radiology-native metric and may miss modality-specific properties. However, FID findings were corroborated by the clinical Turing test.\\

Following FID analysis, scans with gross anatomical abnormalities were manually excluded as a pre-processing step prior to expert validation and public release. This reflects an important limitation to synthetic clinical image generation: the diversity of outputs will inherently include anatomically implausible scans. Furthermore, given the manual nature of review, some scans featuring abnormal anatomy may still be included in our released dataset. Future work should improve DDPM performance to minimize the generation of implausible scans and/or institute automated post-generation filters.\\

Memorization remains a central concern for synthetic medical images. In our top-k nearest neighbor audit using ViT embeddings, two independent reviewers found no visually explicit memorization among the 100 most-similar real-synthetic pairs. While a negative finding in a nearest-neighbor screen indicates that no direct regurgitation of images occurred, it cannot guarantee the absence of memorization. Development of a standardized approach to fidelity, utility, and privacy benchmarking for radiographic imaging data, similar to tabular data \cite{yan2022}, is warranted. This may include analyses featuring stress tests, which vary training set size and sampling parameters.\\

Our dataset and clinical Turing test focused on lateral cervical spine radiographs with a pathology distribution shaped by the source dataset. Generalization to other projections, imaging modalities, and richer pathology remains to be demonstrated. Future experiments should quantify the extent to which DDPMs can retain aberrant/pathologic anatomy found in training data and whether these are preserved at similar prevalence to RWD.\\

\section{Conclusion}

This study demonstrates a growing accuracy and value of synthetic neuroimaging data. Utilizing a DDPM trained on an open-access cervical spine radiographic atlas, we synthesized lateral X-rays that blinded subspecialists rated as indistinguishable from true clinical images. The novel application of DDPM in neuroimaging highlights the potential to generate large-scale imaging datasets to support data-sharing and ML model training for landmarking, segmentation, and classification tasks. 

\newpage

\bibliographystyle{unsrt}  
\bibliography{references}  

\newpage

\section{Supplementary Materials}

\renewcommand{\thetable}{S\arabic{table}}


\begin{table}[h]
    \centering
    \captionsetup{width=0.8\textwidth}
    \caption{Number of Generated and Excluded Synthetic Images Across Model Checkpoints. The table displays the total number of synthetic images generated at three model checkpoints and the number excluded during manual review due to gross anatomical abnormalities. For each checkpoint, an initial set of 1,000 images were generated (‘initial sampling’), from which a subset was used for the blinded validation task. Additional images were generated from each checkpoint (‘large generation’) for the large-scale dataset. Sampling was conducted up to a maximum institutional runtime limit of 24 hours.\\
}
    \label{tab:supplementary_S1}
    \begin{tabular}{lrr}
        \toprule
        Generation & Generated Images & Final Count \\
        \midrule
        Checkpoint 34 (initial sampling) & 1,000 & 607 \\
        Checkpoint 40 (initial sampling) & 1,000 & 654 \\
        Checkpoint 80 (initial sampling) & 1,000 & 742 \\
        Checkpoint 34 (large generation) & 8,000 & 5,071 \\
        Checkpoint 40 (large generation) & 7,968 & 5,999 \\
        Checkpoint 80 (large generation) & 8,032 & 6,990 \\
        \midrule
        Total & 27,000 & 20,063 \\
        \bottomrule
    \end{tabular}
\end{table}

\end{document}